\documentclass[a4paper]{jpconf} 
\usepackage{graphicx}
\usepackage[table]{xcolor}
\usepackage{lipsum}
\usepackage{amsmath}
\usepackage{enumerate}
\usepackage{siunitx}
\usepackage{caption}
\usepackage{subcaption}
\usepackage{array,multirow}
\newcolumntype{C}{ >{\centering\arraybackslash} m{0.4cm} }

\begin{document}
\title{Application of a plasma-based energy booster\\ in operational free-electron lasers}
\author{S Schr\"oder$^{1,2}$, J Osterhoff$^{1,2}$, S Wesch$^1$, C B Schroeder$^{2}$}
\address{$^1$ Deutsches Elektronen-Synchrotron DESY, 22603 Hamburg, Germany}
\address{$^2$ Lawrence Berkeley National Laboratory, Berkeley, California 94720, USA}
\ead{sschroeder@lbl.gov}

\begin{abstract}
Plasma-based accelerators are a promising approach for reducing the size and cost of future particle accelerators, making them a viable technology for constructing and upgrading X-ray free-electron lasers (FELs). Adding an energy booster stage to the linear accelerator of an operational X-ray FEL is recognised as a realistic near-term application of plasma accelerators, with a significant impact on the scientific reach of these facilities. Here, we discuss potential use cases of such a plasma-based energy booster and apply particle-in-cell simulations to estimate its ability to enhance the performance of existing X-ray FEL facilities.
\end{abstract}

\section{Motivation}
Free-electron lasers (FELs)\cite{Motz_JApplPhys_1951, Madey_JApplPhys_1971} stand as unparalleled tools for exploring matter at molecular to atomic scales, providing coherent and ultra-short femtosecond photon pulses. FELs with wavelength extending into the hard X-ray regime ($\mathcal{O}(\SI{0.1}{\angstrom})$) excel as photon sources in pulse brightness, average and peak power, enabling major scientific breakthroughs in structural biology\cite{Chapman_Nature_2011, Seibert_Nature_2011}, atomic physics\cite{Young_Nature_2010}, and material science\cite{Vinko_Nature_2011}, among other fields. Based on slightly different technological concepts and designs, each of today's FEL is a unique facility (see Figure \ref{fig:WavelengthRange}) attracting users with corresponding profiles, yet sharing the common constant pursuit of expanding scientific reach. The performance of an FEL is primarily determined by its radiation wavelength $\lambda_r$  and brilliance $B$.
The radiation wavelength determines the spatial resolution of an FEL:
\begin{equation} \label{eq:wavelength}
\lambda_r \approx \frac{\lambda_u}{2\gamma^2}\left(1+\frac{K^2}{2}\right),
\end{equation}
where $\gamma$ is the relativistic gamma factor, $\lambda_u$ and $K$ are the undulator period and undulator strength, respectively. The quality of the radiation is directly related to the quality of the electron beam and is expressed through the brilliance:
\begin{equation} \label{eq:Brilliance}
B = \left[\frac{\text{ \# photons}}{\text{s}\cdot \text{mrad}^2\cdot \text{mm}^2\cdot0.1\%\text{ BW}}\right] \propto \frac{\langle f \rangle I}{\varepsilon_x\varepsilon_y\delta},
\end{equation}
where the transverse electron beam emittances $\varepsilon_{x/y}$ defines the cross-section and the divergence of the photon pulse, the energy spread $\delta$ determines the radiation bandwidth (BW), the beam current $I$ defines the radiation intensity, and the average repetition rate $\langle f \rangle$ determines the resulting average power of the X-ray beam. The $1/\gamma^2$ scaling of the radiation wavelength but also the pulse pattern (i.e. the pulse repetition rate and the length of the pulse train) depending on the capabilities of the underlying linear electron accelerator (linac) make the linac an integral part of FEL system upgrade considerations. Today's FEL linacs, based on radiofrequency (RF) acceleration technologies, account for 30--50\% of the considerable size of FELs, ranging from hundreds of metres up to several kilometres. A substantial boost in final electron beam energy---embedded into the layout of the existing facility---thus requires compact, high-gradient acceleration. In the following, we discuss the usage of a single plasma accelerator stage based on the external injection beam-driven plasma wakefield acceleration scheme\cite{Ruth_PartAccel_1985, Chen_PRL_1986, Blumenfeld_Nature_2007}.\\

\begin{figure*}[h]
\centering
\includegraphics[trim={25mm 25mm 30mm 25mm},clip]{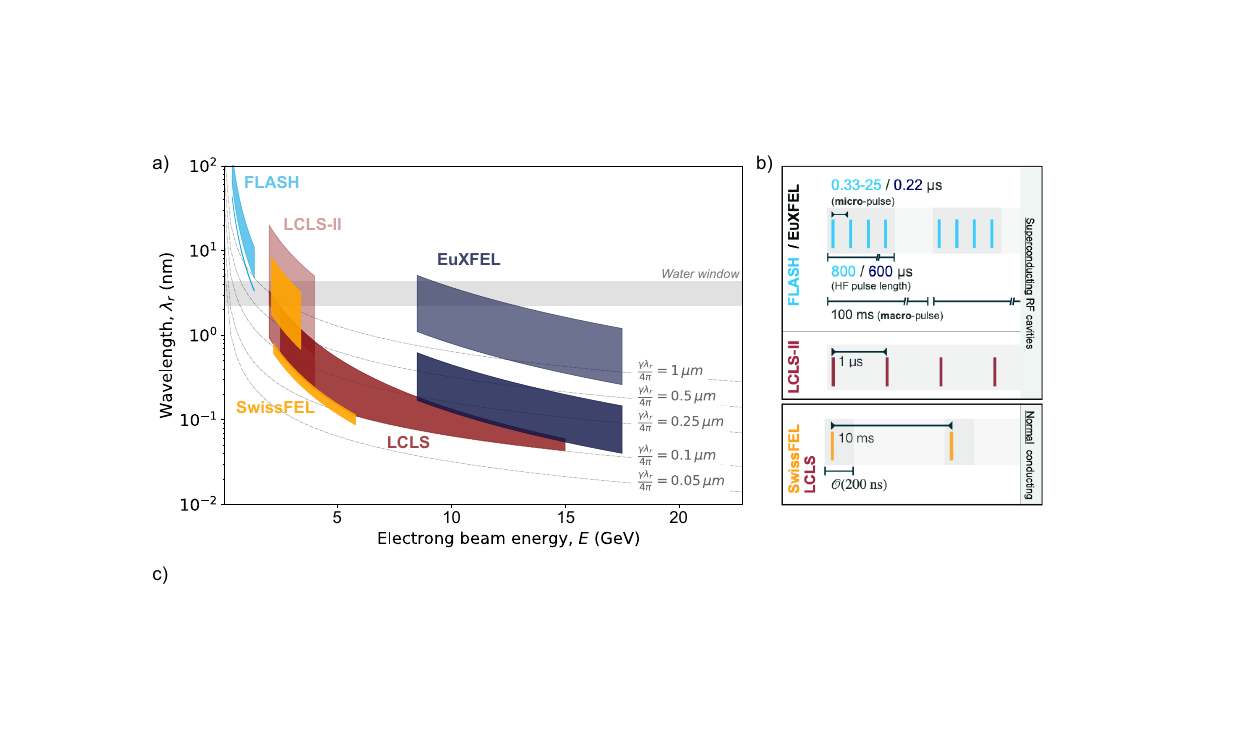}
\label{fig:WavelengthRange}
\vfill
\renewcommand{\arraystretch}{1.6}
\footnotesize
\centering
\begin{tabular}{ | l | c | c | c | c | c | }
\hline
\rowcolor{lightgray} \multirow{2}{*}{ } & \multicolumn{2}{c|}{\textbf{Undulator}} & \multicolumn{2}{c|}{\textbf{Electron bunch}} & \cellcolor{lightgray}\textbf{Linac}\\
\cellcolor{lightgray} & \cellcolor[HTML]{e8e8e6} $\lambda_u$ (mm) & \cellcolor[HTML]{e8e8e6} K & \cellcolor[HTML]{e8e8e6} E (GeV) & \cellcolor[HTML]{e8e8e6} $\varepsilon_{slice}$ (\textmu m)&  \cellcolor{lightgray}\textbf{length (m)}\\
\cline{1-6}
\textbf{EuXFEL} (SASE1\&2) & 35 & 1.65--3.9 & 8--17.5 & 0.1 & \multirow{2}{*}{2000}\\
\textbf{EuXFEL} (SASE3) & 68 & 4--9 & 8--17.5 & 0.1 & \\
\hline
\textbf{FLASH1} & 27.3 & 1.17--1.23 & 0.38--1.35 & 0.5 & \multirow{2}{*}{135}\\
\textbf{FLASH2} & 31.4 & 1.5--2.8 & 0.38--1.35 & 0.5 & \\
\hline
\textbf{LCLS-II} (1) & 39 & 1.24--5.43 & 4 & 0.2 & \multirow{2}{*}{930}\\
\textbf{LCLS-II} (2) & 26 & 0.44--2.44 & 4 & 0.2 & \\
\hline
\textbf{LCLS} & 30 & 3.5--3.6 & 4 & 0.1 & 650\\
\hline
\textbf{SwissFEL} (Athos) & 40 & 1--3.05 & 2.1--3.4 & 0.1 & 290\\
\textbf{SwissFEL} (Aramis) & 15 & 1--1.04 & 2.1--5.8 & 0.1 & 440\\
\hline
\end{tabular}
\vspace{0.7cm}
\caption{\textbf{FEL facility specifications} (a) Achievable radiation wavelength range according to Equation \ref{eq:wavelength}. FLASH, SwissFEL, LCLS-II and EuXFEL comprise two beamlines equipped with different undulator types and/or operated at different energies (SwissFEL) covering different wavelength ranges. (b) Pulse-train timing pattern of facilities. (c) Undulator and electron bunch specifications of facilities \cite{SLAC_machine_parameters, FLASH_machine_parameters, XFEL_machine_parameters, SwissFEL_machine_parameters}.}
\label{fig:WavelengthRange}
\end{figure*}

Plasma wakefields sustain electric fields on the order of gigavolt-per-meter (GV/m)\cite{TajimaDawson_PRL_1979}, enabling the acceleration of electrons with orders of magnitude higher acceleration gradients than today's RF technology (MeV/m). Amplified undulator radiation originating from an electron beam accelerated in a \textit{laser-driven} plasma wakefield \cite{Wang_Nature_2021, Labat_NaturePhotonics_2023} recently represented a major milestone in the field of plasma accelerators. While the laser-driven scheme promises the realisation of truly compact linacs, the high-power lasers currently available for wakefield excitation are orders of magnitude away from supporting the megahertz repetition rate of today's cutting-edge FELs and thus meet the user profiles only for low-power operation. \textit{Beam-driven} wakefield excitation, on the other hand, is an interesting way also to facilitate high-power operation. Based on continuous development and the establishment of novel methodologies for operating a beam-driven plasma wakefield accelerator\cite{Rosenzweig_PRL_1988, Muggli_IEEE_1999, Muggli_PRL_2004, Hogan_PRL_2005, Schroeder_JPhysConfSer_2020, Lindstrom_PRAB_2020, Schroeder_NatCommun_2020}, significant progress has been made over the past decade towards demonstrating high-efficiency acceleration \cite{Litos_Nature_2014}, beam quality preservation\cite{Lindstrom_PRL_2021, Lindstrom_ResearchSquare_2022}, and first amplified undulator radiation \cite{Pompili_Nature_2022}. Achieving high average power remains a critical milestone to confirm applicability as an energy booster in FELs at full capacity, with the acceleration of bunch trains at megahertz inter-bunch repetition rate while maintaining beam characteristics\cite{Loisch_EAAC_2023} already presented.

\begin{figure*}[h]
\includegraphics[trim={25mm 30mm 25mm 35mm}, width=\textwidth]{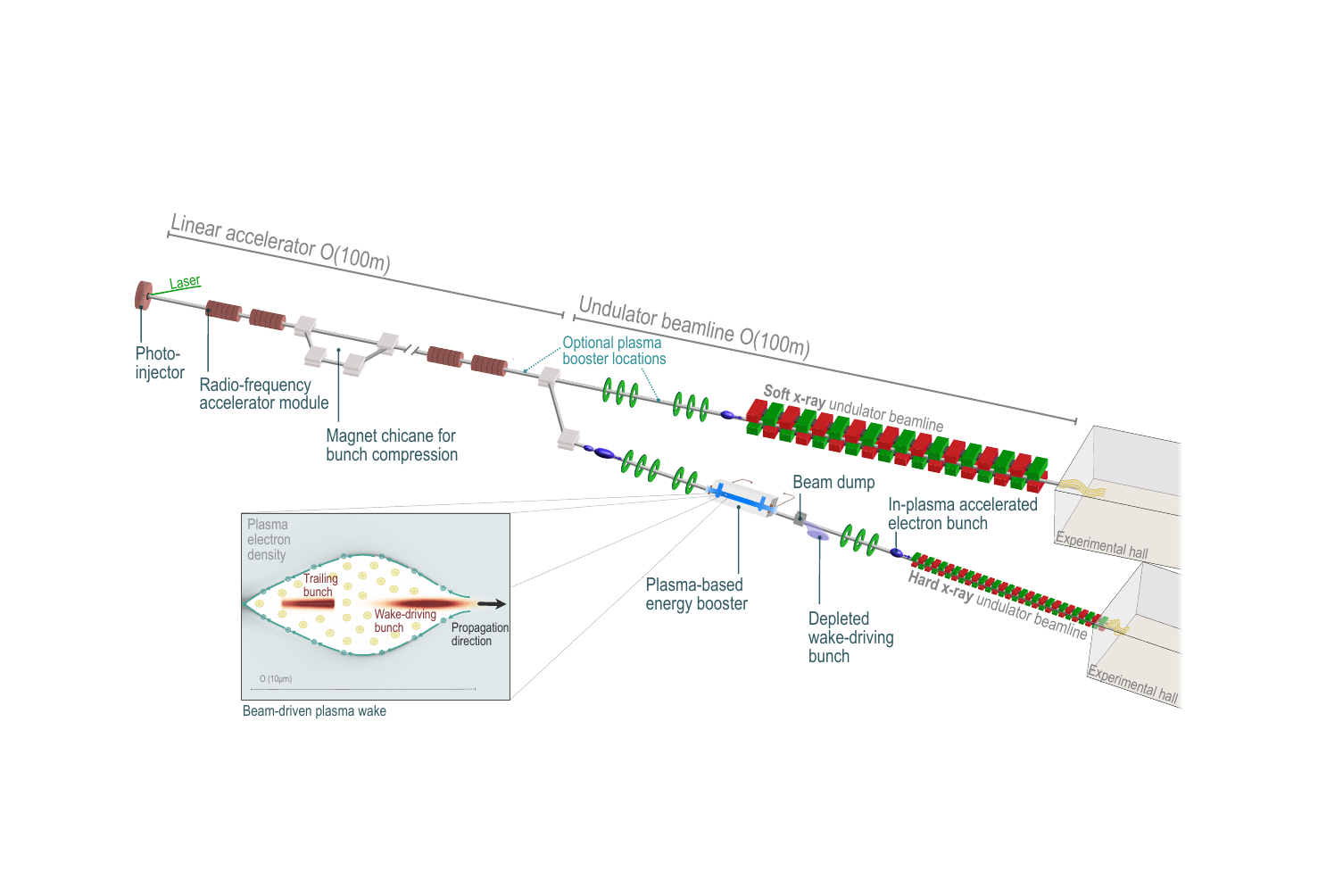}
\caption{\textbf{Exemplary plasma booster scheme}. The linear accelerator of an FEL is extended by a plasma accelerator stage. Two consecutive bunches are accelerated in the linac. Tightly focused into a plasma channel, the first bunch drives a cavity-like wake structure with pronounced charge separation of the mobile plasma electrons and the heavier, immobile plasma ions. The resulting longitudinal and transverse electromagnetic fields accelerate and focus the trailing bunch. The energy-depleted driving bunch is extracted from the beamline; the accelerated trailing bunch is further delivered to the undulator section for X-ray generation. Depending on the intended purpose, the plasma stage can be integrated into individual beamlines, allowing the operation of the beamlines at independent beam energies (as shown here), or alternatively, it could serve as a common energy booster after the linac.}
\label{fig:setup}
\end{figure*}

\section{Plasma booster scheme}
\label{sec:BoosterScheme}
In the scheme of an external injection beam-driven plasma wakefield electron energy booster stage for existing FELs (see Figure \ref{fig:setup}), both the wake-driving and the trailing to-be-accelerated electron bunch are provided by the existing FEL linac. With an FEL linac typically consisting of a photo-cathode followed by RF-based acceleration sections interleaved with magnetic chicanes for longitudinal bunch compression, the low-emittance electron bunches exhibit a typical kilo-Ampère peak current with femtosecond duration---making them ideally suited for resonantly driving strong wakefields in a plasma. The plasma accelerator stage then acts as an energy transformer, wherein the wake-driving bunch transfers its energy to the trailing bunch, with the plasma acting as the mediating medium. After the acceleration process in the plasma, the energy-depleted bunch that has driven the wakefield is extracted from the beam transfer line, while the accelerated trailing bunch is transferred to the undulator section for radiation generation. Such an energy booster can fulfil various targeted linac upgrade objectives, which can be categorised as follows:
\begin{enumerate}
    \item \textit{Reaching shorter X-Ray wavelength}. According to Equation \ref{eq:wavelength}, an energy boost generally enables shorter photon wavelengths. Whilst every FEL would, in principle, benefit from a shorter wavelength range, i.e., a higher spatial resolution, not every facility is equally suited to achieve this through an electron energy boost. High-quality radiation generation requires low electron beam emittance $\varepsilon_n<(\gamma \lambda_r) / (4 \pi) \propto 1/\gamma$ (dotted lines in Figure \ref{fig:WavelengthRange}a) and energy spread $\sigma_{\delta} < \rho \propto 1/\gamma$ with the Pierce parameter $\rho$. Fulfilling both requirements becomes increasingly challenging or even unfeasible with an energy boost. As it stands today, a plasma booster could, in particular, help FLASH cover the water window fully (see Figure \ref{fig:WavelengthRange}a)--the most relevant wavelength range for biological studies, or help LCLS-II to realise the already planned electron energy doubling in a more cost-effective plasma-based variant. In contrast, the wavelength range of LCLS and SwissFEL with today's configuration is already limited by beam emittance and energy spread, respectively.
    \item \textit{Parallel operation of beamlines at independent beam energies}. Due to the shared linear accelerator in FELs comprising more than one beamline, the beam energy and, consequently, the achievable wavelength range in different experimental areas are interdependent. Implementing a plasma booster in the individual beamlines would allow their operation at different energies. For instance, the soft X-ray EuXFEL beamline could operate in the water window, and the hard X-ray beamline still reaches the lowest achievable wavelength simultaneously. With a plasma booster, FELs that are typically overbooked gain operational flexibility; costly downtime of individual beamlines due to incompatibilities between user experiments will be reduced.
    \item \textit{Increasing pulse train length.} The timing pattern of the photon pulse train (see Figure \ref{fig:WavelengthRange}b) dictates the observable dynamics. In a simplified picture, the pulse width establishes the temporal resolution, the temporal gap between pulses defines the sampling rate, and the length of the pulse train determines the maximum observable duration of a dynamics. Thus, there is a pursuit of ever shorter pulses, higher pulse-to-pulse repetition rates, and long pulse trains. However, this has technological limits, namely the power acceptance in the accelerator modules. Linear accelerators based on superconducting radio-frequency (SRF) accelerator technology, operated at the maximum achievable acceleration gradient, require a low duty cycle for cooling, leading to a characteristic macro-/micro-pulse structure (FLASH, EuXFEL) (see Figure \ref{fig:WavelengthRange}b). Continuous pulse operation using SRF technology (LCLS-II) requires a reduction in the acceleration gradient and, consequently, a larger facility or operation at lower electron energy. Employing a plasma booster would allow an increase of the duty cycle in SRF modules by reducing the acceleration gradient without impacting the final beam energy. A favourable photon pulse train pattern could be achieved without compromising the attainable wavelength range.
\end{enumerate}

Beyond these FEL upgrade applications, electron bunches with $\mathcal{O}(\SI{10}{GeV})$ energies are imperative for future strong-QED\cite{Abramowicz_EurPhysJSpecTop_2021, Turner_EurPhysJD_2022} and photo-nuclear\cite{Zilges_2022} experiments. A plasma booster stage could significantly increase the scientific reach of those studies conducted in FELs, while the lower beam quality and average power requirements serve the booster as a stepping stone to a full-capacity plasma booster for actual FEL operation. Likewise, injector linacs of synchrotrons and storage rings could employ a plasma booster.\\

Despite keen interest from the plasma accelerator as well as the FEL community, there is currently no estimate for the achievable charge and energy of a plasma booster system in existing free-electron lasers. The development of such a plasma-based energy booster hinges on its intended application and even more on the available infrastructure. In contrast to a detailed design study tailored to the individual FEL facilities, the aspiration of this study is, firstly, to provide an overview of the wealth of possible applications of a booster stage and, secondly, to introduce quantitative figures into the discussion---following the philosophy of keeping the existing facilities unchanged as much as possible, exploiting the idealised capabilities of a plasma accelerator stage.

\section{Methodology}
The achievable acceleration is fundamentally limited by the transferable energy of the wake-driving bunch, $E_{kin}Q$, with the kinetic energy $E_{kin}$ and the charge  $Q$. To estimate a physical upper limit, we fully exploit the currently available electron bunch at the facilities for wakefield excitation and assume that the trailing bunch can be generated and accelerated in the same linac\footnote{This assumption on two-bunch generation is non-trivial. Testing its feasibility requires simulations that consider collective effects (space charge, coherent synchrotron radiation), and complementing proof-of-concept experiments.}. The plasma response is simulated with single-step particle-in-cell (PIC) simulations, based on which the achievable acceleration gradient and bunch charge are estimated. The methodology of the simulation study is illustrated in Figure \ref{fig:Methodology} and follows these steps:

\begin{figure}[ht]
\begin{minipage}{.5\textwidth}
\centering
\includegraphics[trim={30mm 25mm 27mm 30mm},width=\textwidth]{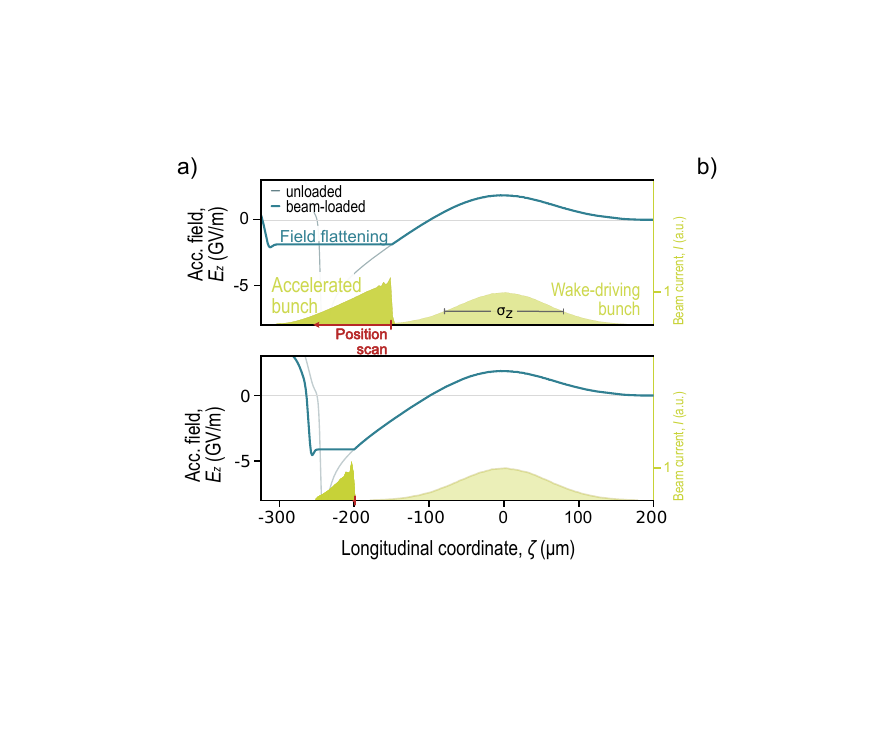}
\end{minipage}
\hfill
\begin{minipage}[t]{0.5\textwidth}
\footnotesize
\centering
\renewcommand{\arraystretch}{2}
\vspace{-3.cm}
\begin{tabular}{  | l | C C C C C | }
\hline
\rowcolor{lightgray} & E & Q & $I_{p}$ & $\sigma_z$& $\varepsilon_{x/y}$\\
\rowcolor{lightgray} & \cellcolor[HTML]{e8e8e6}\tiny{(GeV)} & \cellcolor[HTML]{e8e8e6}\tiny{(nC)} & \cellcolor[HTML]{e8e8e6}\tiny{(kA)} & \cellcolor[HTML]{e8e8e6}\tiny{(\textmu m)} & \cellcolor[HTML]{e8e8e6}\tiny{(\textmu m)}\\
\hline
\textbf{FLASH} & 1 & 1 & 2 & 60 & 0.5\\ \hline
\textbf{EuXFEL} & 17.5 & 1 & 6 & 30 & 0.1\\ \hline
\textbf{SwissFEL} & 3.4 & 0.2 & 0.8 & 12 & 0.1\\ \hline
\textbf{LCLS-II} & 4 & 0.1 & 0.8 & 10 & 0.1\\
\hline
\end{tabular}
\end{minipage}
\caption{\textbf{Methodology of simulation study.} (a) Two consecutive bunches, the wakefield-driving and to-be-accelerated trailing bunch, are tightly focused into a plasma. The first bunch drives a nonlinear wakefield in which the trailing bunch is accelerated. The excited wakefield (dotted grey line) can be flattened over the extent of the trailing bunch (solid blue line) by means of beam loading when precisely tailoring the current profile of the trailing bunch. In order to change the acceleration gradient, the position of the trailing bunch is scanned over the accelerating wake phases (see bottom panel), which also changes the loadable bunch charge. (b) Beam parameters of the wakefield-driving bunch.}
\label{fig:Methodology}
\end{figure}

\begin{enumerate}
\item \textit{Drive-bunch generation with today's available electron bunch parameters.} Lowest bunch emittance, nominal correlated energy spread, maximum available bunch charge, maximum achievable bunch energy, and a Gaussian current profile with 80\% of the maximum achievable peak current are assumed for the wake-driving bunch (see Figure \ref{fig:Methodology}c). Maximum charge and energy allow driving and sustaining the acceleration over the longest distance with maximum power and the highest beam rigidity. The limitation of the peak current $I_p$ to 80\% makes the assumption of a Gaussian current profile more realistic and mitigates coherent synchrotron radiation induced instabilities at maximum compression.
\item \textit{Longitudinal and transverse matching.} For resonant excitation of the wakefield, the wake-driving bunch's length must match the resulting wakefield's decelerating phase. To the peak current at its maximum, the plasma density is chosen according to the bunch length in the range of $k_p\sigma_z = [0.6, 1.8]$, where $k_p^{-1} = c/\omega_p$ is the plasma skin depth with $c$ the speed of light and $\omega_p = \sqrt{4\pi n_p e^2/m_e}$ (in CGS units) the plasma frequency with the elementary electron charge $e$ and electron mass $m_e$. Higher plasma density will generally allow higher gradients to be achieved, but require smaller bunch dimensions and reduce the maximum charge load. Both the wake-driving and trailing bunch are transversely matched to the focusing fields in the plasma, allowing for in-principle transverse beam-quality preservation.
\item \textit{Current profile shaping of the trailing bunch for field flattening.} Injecting charge into the plasma wakefield changes its shape by means of beam loading\cite{Katsouleas_PRA_1986}. This effect fundamentally limits the charge that can be injected into a plasma accelerator. By precisely shaping the current profile of the injected charge, beam loading can be used to flatten the field over the wake phase extent of the to-be-accelerated bunch (see Figure \ref{fig:Methodology}a) such that the bunch is uniformly accelerated and does not exhibit energy spread during acceleration\cite{Katsouleas_PRA_1986, Tzoufras_PRL_2008}---a necessity for scalable acceleration over long plasma distances and loss-free bunch transport.
\item \textit{Scanning the acceleration gradient via the trailing bunch position}. The acceleration gradient is determined by the wakefield phase at which the trailing bunch is injected, which at the same time determines the maximum loadable charge under the assumption of field flattening. The link between achievable gradient and loadable charge is gained by scanning the position of the trailing bunch and tailoring the current profile at each injection position (compare top and bottom panels in \ref{fig:Methodology}a).
\end{enumerate}

The simulations are conducted with the 3D quasi-static PIC code HiPACE++\cite{Diederichs_CPC_2022} using the built-in SALAME algorithm for current-profile shaping\cite{{Diederichs_PRAB_2020}}.
Both bunches were modelled with $10^6$ numerical particles, with the macro-particles of the drive bunch having constant weight. The plasma was sampled with 4 particles per cell. A simulation box of size $20 \times 20 \times 16 $ $k_p^{-3}$ (in $x \times y \times \zeta$) was resolved by a grid of $1024 \times 1024 \times 600$ cells.

\section{Results and discussion}
\begin{figure*}
\includegraphics[trim={25mm 30mm 25mm 25mm}, width=\textwidth]{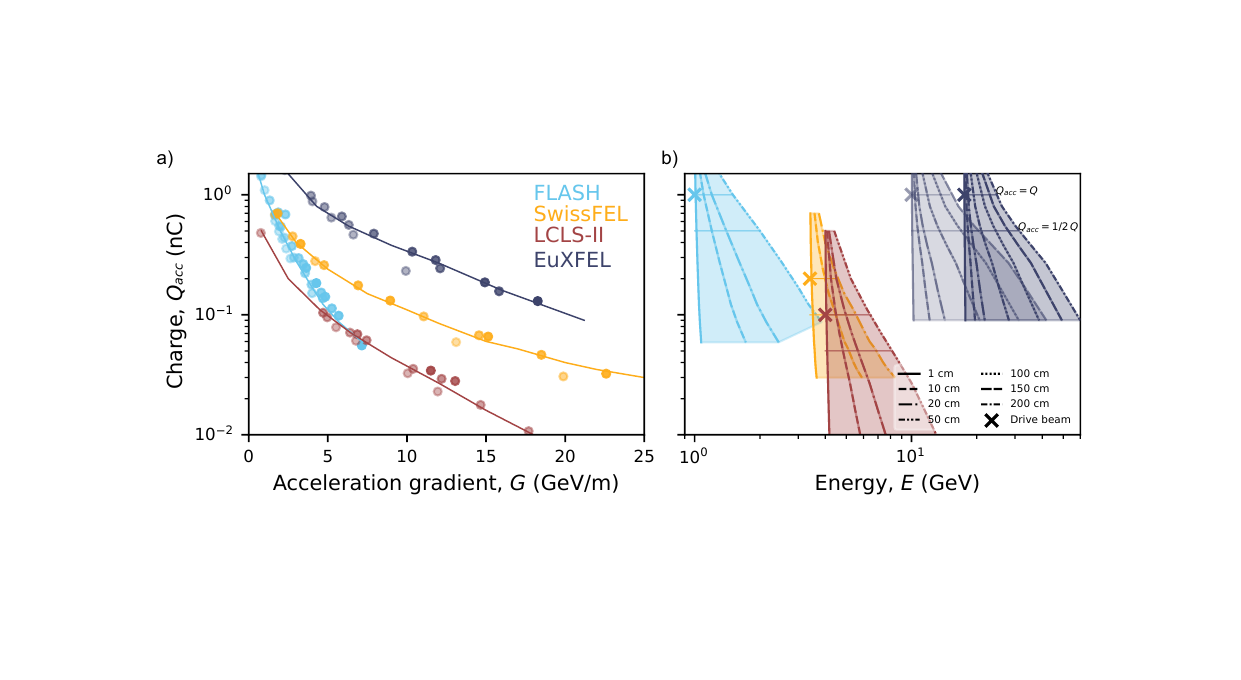}
\caption{\textbf{Simulation results.} (a) Coupling between maximum loadable charge and achievable acceleration gradient under the assumption of optimal beam loading. (b) Resulting bunch charge and energy for variable plasma channel lengths. Two horizontal lines across each facility phase space highlight scenarios where the injected bunch has the same and half the charge as the initial bunch at the facilities. For the EuXFEL two options are shown with either an initial beam energy of 17\,GeV or 10\,GeV (lighter color).}
\label{fig:GradientChargeDependency}
\end{figure*}

Based on the discussion about the most potent applicability of a plasma booster (Section \ref{sec:BoosterScheme}), simulations are performed for FLASH, SwissFEL, LCLS-II, and EuXFEL. The expected trend of a lower loadable charge, $Q_{acc}$, at higher acceleration gradients, $G$, is found across all facilities (see Figure \ref{fig:GradientChargeDependency}a); the rate varies. The rapid reduction in bunch charge with increased acceleration gradient for FLASH can be attributed to the comparatively long bunch length and the correspondingly lower plasma densities. The loadable charge depending on the acceleration gradient is empirically estimated (solid lines) to extrapolate the achievable beam energy in finite plasma channel lengths from 1\,cm up to 2\,m (see Figure \ref{fig:GradientChargeDependency}b). For beam quality preservation, the driven wakefield must not change over the acceleration length. The maximum acceleration length is given by the depletion length, $L_{dep}=E_{drive}/E_{z, max}$, at which the part of the drive-bunch that experiences the maximum decelerating field ($E_{z, max}$) has transferred its energy, $E_{drive}$, to the plasma. At the EuXFEL, the acceleration could be sustained over meter-long plasma channels owing to the high initial beam energy, whereas at LCLS-II and FLASH the acceleration is limited to 50\,cm plasma channel length and at SwissFEL to 20\,cm. In the case of no charge reduction of the accelerated beam, an energy gain of 0.5\,GeV, 1.2\,GeV, 2.5\,GeV, 7\,GeV to a final beam energy of 1.85\,GeV, 3.3\,GeV, 6.5\,GeV, 24.5\,GeV can be achieved for FLASH, SwissFEL, LCLS-II, and EuxFEL, respectively. If the final bunch charge can be reduced to 50\%, an increased energy gain of 1\,GeV, 2\,GeV, 4\,GeV, 14\,GeV could be achieved for FLASH, SwissFEL, LCLS-II, and EuxFEL, respectively---representing energy-doubling for FLASH and LCLS-II in only 50\,cm. Operating EuXFEL at reduced initial energy of 10\,GeV, an energy gain of 5\,GeV and 11\,GeV can be achieved for equal and 50\% reduced bunch charge, respectively.

This study is based on fundamental assumptions that need to be revisited in the context of these results. First, the available bunches at the facilities were used unchanged for driving the wakefield. Second, a perfect flattening of the field was required. When assuming the drive-bunch current profile to be also tailored, the transformer ratio, $E_{z, acc}/E_{z, max}$, \cite{Loisch_PRL_2018, Roussel_PRL_2020} and depletion length increases, i.e., the energy transfer efficiency enhances\cite{Pena_2023}. The results presented here can therefore be considered conservative estimates, and further optimisation is possible. Depending on the scheme, perfect field flattening for energy-spread preservation may not be necessary. For attosecond-pulse schemes\cite{Emma_APLPhoton_2021}, an increased correlated energy spread may be desirable for post-plasma compression. Or, the typically negative chirp of the initial beam can be reduced via overloading the wakefield. However, owing to the inherently large changes in wakefield strength over the small scale of wake structure, the current profile cannot deviate significantly from the here optimised cases if an acceleration on the meter scale is to be sustained. The values presented here are therefore also relevant for advanced schemes.

The here shown trade-off between bunch charge and final beam energy is used as a figure of merit to infer on the trade-off between radiation brilliance and wavelength. Using the bunch charge as the main indicator for the impairment of brilliance is not entirely correct, as the current profile is ultimately decisive. The low-current tail of the trailing bunch may not contribute to the radiation generation in the undulators. Given the consistently triangular shape and the limited extent of the wake structure, however, the average current of the bunch scales with its bunch charge and is therefore a valid measure of the qualitative effect on the resulting brilliance: lower charge results in lower brilliance. Furthermore, as discussed above, FELs have stringent requirements not only on the energy spread for the generation of high-quality amplified undulator radiation, but also on the beam emittance. Although emittance preservation is theoretically possible in the blowout regime, it is technically increasingly difficult to achieve over a long plasma channel where any transverse misalignment or source of instability grows. Given the requirements of both parameters, energy spread and emittance, scaling with $1/\gamma$, the ultimately achievable FEL radiation with the accelerated bunches discussed here must be thoroughly quantified with simulations and tolerance studies must be performed that consider the transverse evolution of the bunches in the plasma and model the resulting radiation process in the undulators.

\section{Conclusion and Outlook}
Beam-driven plasma wakefield accelerators have made significant strides in attaining electron beam characteristics relevant to X-ray FELs in recent years. Consequently, exploring the quantitative reach of a plasma-based accelerator stage as an upgrade option for existing free-electron lasers is pertinent. 

In the context of an all-encompassing upgrade of an FEL, equal attention must be paid to the wavelength and brilliance of the radiation. A plasma booster that upgrades an FEL to its full capacity must support the operation with its diverse user groups with individual requirements for both performance parameters and, therefore, should not compromise one against the other. However, individual users may benefit from a significantly extended wavelength range at lower brilliance and vice versa. The simulation results here highlight a wide range of highly pertinent possibilities for upgrading FELs through a plasma booster, which can greatly increase the phase space of the available bunch charge and energy and, thus, the flexibility of the facility. The study here focused on the limitations that arise through beam loading. To gain a complete picture of the ultimate limitations that also comprise effects such as head erosion, ion motion, and hosing, full-3D PIC simulations along the full acceleration length with evolving bunches are required. Beyond pure theoretical limitations, to also understand practical limitations, an extended tolerance study based on start-to-end simulations is required including collective effects and the FEL process. It is worth noting, that the gain length scales linearly with beam energy, reaching similar saturation with higher beam energy will thus also require an upgrade of the undulator beamline and requires further studies.

Despite the ongoing successes in external injection beam-driven plasma wakefield acceleration in recent years, implementing such an energy booster stage still demands significant R\&D efforts and encounters technical hurdles. First and foremost, enabling high average power operation is imperative. The precise two-bunch generation already represents a vast area of exploration and is one of the most crucial technological developments to be made towards the realisation of a booster stage. Last, but not least, the routine operation of these micrometre-scale plasma accelerator structures demands precision in beam delivery for both in- and out-coupling, surpassing the capabilities of today's FELs. Novel approaches employing computer-aided, autonomous precision control and electron beam diagnostics will be indispensable and may revolutionise FEL operation itself. The discussed booster stage is not only relevant for upgrading existing FELs but is also an attractive avenue for constructing additional FELs worldwide and preparing FELs for industrial applications. A specialised service centre to support drug development through individualised machines for the targeted development of advanced materials could be a possible industrial application. Advanced operational controls will hereby play a pivotal role in reducing operational costs—especially personnel expenses, which currently constitute about 50\% of the operational costs, with a need for highly specialised experts available only at a few major research laboratories around the world. 

The successful integration of a plasma-based electron energy booster into the operation of an FEL will represent a milestone in plasma wakefield acceleration and demonstrate the maturity of this technology on the path to future high-energy physics applications.

\ack
C.B.S, J.O., and S.S. are supported by the Director, Office of Science, Office of High Energy Physics, of the U.S. Department of Energy under Contract No. DE-AC02-05CH11231. S.S. was supported by the Helmholtz Innovation and Transfer Center via the Field Study Fellowship.


\section*{References}
\begin{thebibliography}{9}
\bibitem{Motz_JApplPhys_1951} Motz H 1951 \textit{J. Appl. Phys.} \textbf{22} 527
\bibitem{Madey_JApplPhys_1971} Madey J M J 1971 \textit{J. Appl. Phys.} \textbf{42} 1906
\bibitem{Chapman_Nature_2011} Chapman H \textit{et al.} 2011 \textit{Nature} \textbf{470} 73-78
\bibitem{Seibert_Nature_2011} Seibert M M \textit{et al.} 2011 \textit{Nature} \textbf{470} 78-82
\bibitem{Young_Nature_2010} Young  L \textit{et al.} 2010 \textit{Nature}, \textbf{466} 46-52
\bibitem{Vinko_Nature_2011} Vinko S M \textit{et al.} 2011 \textit{Nature} \textbf{482} 59-62
\bibitem{Ruth_PartAccel_1985} Ruth R \textit{et al.} 1985 \textit{ Part. Accel.} \textbf{17} 171
\bibitem{Chen_PRL_1986} Chen P \textit{et al.} 1986 \textit{Phys. Rev. Lett.} \textbf{56} 1252
\bibitem{Blumenfeld_Nature_2007} Blumenfeld I \textit{et al.} 2007 \textit{Nature} \textbf{445} 741–744
\bibitem{SLAC_machine_parameters} https://lcls.slac.stanford.edu/machine/parameters, accessed 31.12.2023 12:14
\bibitem{FLASH_machine_parameters} https://flash.desy.de/accelerator/, accessed 31.12.2023 13:11
\bibitem{XFEL_machine_parameters} Altarelli M \textit{et al.} 2007 The European X-ray Free-Electron Laser Technical design report
\bibitem{SwissFEL_machine_parameters} Ganter R \textit{et al.} 2010 SwissFEL Conceptual Design Report 
\bibitem{TajimaDawson_PRL_1979} Tajima T and Dawson J M 1979 \textit{Phys. Rev. Lett.} \textbf{43} 267
\bibitem{Wang_Nature_2021} Wang W \textit{et al.} 2021 \textit{Nature} \textbf{595} 516–520 
\bibitem{Labat_NaturePhotonics_2023} Labat M \textit{et al.} 2023 \textit{Nat. Photonics} \textbf{17}, 150–156
\bibitem{Rosenzweig_PRL_1988} Rosenzweig JB \textit{et al.} 1988 \textit{Phys. Rev. Lett.} \textbf{61}, 98
\bibitem{Muggli_IEEE_1999} Muggli P \textit{et al.} 1999 IEEE Trans. (Plasma Sci.) 27, 791–799
\bibitem{Muggli_PRL_2004} Muggli P \textit{et al.} 2004 \textit{Phys. Rev. Lett.} \textbf{93}, 014802
\bibitem{Hogan_PRL_2005} Hogan M \textit{et al.} 2005 \textit{Phys. Rev. Lett.} \textbf{95}, 054802
\bibitem{Schroeder_JPhysConfSer_2020} Schröder S \textit{et al.} 2020 \textit{J. Phys.: Conf. Ser.} \textbf{1596} 012002
\bibitem{Lindstrom_PRAB_2020} Linstrøm C A \textit{et al.} 2020 \textit{Phys. Rev. Accel. Beams} \textbf{23}, 052802
\bibitem{Schroeder_NatCommun_2020} Schröder S \textit{et al.} 2020 \textit{Nat. Commun.} \textbf{11}, 5984
\bibitem{Litos_Nature_2014} Litos M \textit{et al.} 2014 \textit{Nature} \textbf{515} 92-95
\bibitem{Lindstrom_PRL_2021} Linstrøm C A \textit{et al.} 2021 \textit{Phys. Rev. Lett.} \textbf{126} 014801
\bibitem{Lindstrom_ResearchSquare_2022} Lindstrøm C A \textit{et al.} 2022 (perprint on Research Square)
\bibitem{Pompili_Nature_2022} Pompili R \textit{et al.} 2022 \textit{Nature} \textbf{605}, 659–662
\bibitem{Loisch_EAAC_2023} Loisch G \textit{et al.} EAAC contribution 297 (to be published)
\bibitem{Abramowicz_EurPhysJSpecTop_2021} Abramowicz H \textit{et al.} 2021 \textit{Eur. Phys. J. Spec. Top.} \textbf{230} 2445–2560
\bibitem{Turner_EurPhysJD_2022} Turner M \textit{et al.} 2022 \textit{Eur. Phys. J. D} \textbf{76}:205
\bibitem{Zilges_2022} Zilges A \textit{et al.} 2022 Progress in Particle and Nuclear Physics \textbf{122}, 103903
\bibitem{Katsouleas_PRA_1986} Katsouleas T 1986 \textit{Phys. Rev. A} \textbf{33}, 2056
\bibitem{Tzoufras_PRL_2008} Tzoufras M \textit{et al.} 2008 \textit{Phys. Rev. Lett.} \textbf{101}, 145002
\bibitem{Diederichs_CPC_2022} Diederichs S \textit{et al.} 2022 \textit{Comput. Phys. Commun.} \textbf{278} 108421
\bibitem{Diederichs_PRAB_2020} Diederichs S \textit{et al.} 2020 \textit{Phys. Rev. Accel. Beams} \textbf{23} 121301
\bibitem{Loisch_PRL_2018} Loisch G \textit{et al.} 2018 \textit{Phys. Rev. Lett.} \textbf{121} 064801
\bibitem{Roussel_PRL_2020} Roussel R \textit{et al.} 2020  \textit{Phys. Rev. Lett.} \textbf{124} 044802
\bibitem{Pena_2023} Peña F \textit{et al.} 2023 (preprint arXiv:2305.09581v2)
\bibitem{Emma_APLPhoton_2021} Emma C \textit{et al.} 2021 \textit{APL Photon} \textbf{6} 076107
\end{thebibliography}
\end{document}